\documentclass[twocolumn,showpacs,pra,aps,superscriptaddress,floatfix]{revtex4-1}
\pdfoutput=1
\usepackage{graphicx}
\usepackage{braket} 
\usepackage{float}  				    
\usepackage{natbib}
\usepackage[utf8]{inputenc}
\usepackage{dsfont}
\usepackage{bm}
\usepackage{amsmath}

\usepackage{color}
 \usepackage{soul}
 \usepackage{amsfonts}

 \definecolor{darkred}{rgb}{0.8,0.1,0.1}
 \definecolor{DARKRED}{rgb}{0.8,0.1,0.1}
 \definecolor{darkblue}{rgb}{0.1,0.1,0.7}
 \definecolor{bleudefrance}{rgb}{0.19, 0.55, 0.91}
 
 \DeclareMathOperator{\Tr}{Tr}

\newtheorem{definition}{Definition}
\newtheorem{theorem}{Theorem}

\usepackage{colordvi}

\definecolor{orange}{RGB}{255,127,0}

\begin{document}

\title{Entangling Power of Symmetric Two-Qubit Quantum Gates}

\author{D. Morachis Galindo}
\affiliation{Departamento de Física, Centro de Nanociencias y Nanotecnolog\'ia, Universidad
Nacional Aut\'onoma de M\'exico, Apartado Postal 14, 22800 Ensenada, B.C., M\'exico}%

\author{Jes\'us A. Maytorena}
\affiliation{Departamento de Física, Centro de Nanociencias y Nanotecnolog\'ia, Universidad
Nacional Aut\'onoma de M\'exico, Apartado Postal 14, 22800 Ensenada, B.C., M\'exico}

\date{\today}

\begin{abstract}
The capacity of a quantum gate to produce entangled states on a bipartite system is quantified in terms of the entangling power. This quantity is defined as the average of the linear entropy of entanglement of the states produced after applying a quantum gate over the whole set of separable states. Here we  focus on symmetric two-qubit quantum gates, acting on the symmetric two-qubit space, and calculate the entangling power in terms of the appropriate local-invariant. A geometric description of the local equivalence classes of gates is given in terms of the $\mathfrak{su}(3)$ Lie algebra root vectors. These vectors define a primitive cell with hexagonal symmetry on a plane, and through the Weyl group the minimum area on the plane containing the whole set of locally equivalent quantum gates is identified. We give conditions to determine when a given quantum gate produces maximally entangled states from separable ones (perfect entanglers). We found that these gates correspond to one fourth of the whole set of locally-distinct quantum gates. The theory developed here is applicable to three-level systems in general, where the non-locality of a quantum gate is related to its capacity to perform non-rigid transformations on the Majorana constellation of a state. The results are illustrated by an anisotropic Heisenberg model, the Lipkin-Meshkov-Glick model, and two coupled quantized oscillators with cross-Kerr interaction. 
\end{abstract}
\maketitle

\title{UhlmannJ}
\author{dmorachisgalindo }
\date{February 2021}


\maketitle

\section{Introduction}

Entanglement is a purely quantum mechanical phenomenon that is essential to achieve universal quantum computation based on interacting qubits systems\cite{EntangledSystems}. Quantum logic gates are the building blocks to perform quantum algorithms, where the generation of entangled states from a separable set of states is mandatory to achieve the desired results\cite{QuantumEntanglementHorodecki}.

Most of the proposed quantum computer architectures are based on multi-qubit processors. Nevertheless, there are also proposals that use higher dimensional systems called qudits, which have the advantage of reducing the number of physical entities required to perform calculations\cite{MultivaluedLogicGates}. Among these are three-level systems, called qutrits, which are the smallest systems that may exhibit purely quantum correlations such as contextuality\cite{Contextualitywithoutnonlocality}, and they have been used to construct three-level quantum gates\cite{Arvind,MultilevelSU(2)}. Qutrits may be emulated by a two-qubit system symmetric under particle exchange. This allows us to think of many three-level transformations in terms of operations on the symmetric two-qubit symmetric space.

Within the Majorana representation 
\cite{MajoranaAtomi,MajoranaMultiqubit}, symmetric two-qubit states appear as two points (``the stars'') on the unit sphere.
It can be shown that the distance between the two stars maps bijectively to the concurrence\cite{LiuBerryEntanglement}. States with maximally separated stars correspond to Bell states, while states with stars at the same position are separable. This allows to think of any transformation between this kind of states as rigid or non-rigid motions of the associated Majorana constellation, where the latter (former) does (does not) change the entanglement of states. Whenever there is no place for confusion, we will refer to a two-qubit symmetric state (space) as symmetric state (space) only.


Since many transformations on symmetric (three-level) spaces involve changing the entanglement (distance between the Majorana stars), it is important to quantify the capacity of quantum gates to generate it. Such a quantity is called the \textit{entangling power}\cite{ZanardiEntanglingPOwer}. It is defined as the average linear entropy of the states produced by a quantum gate $\hat{V}$ acting on the manifold of all separable states. For general two-qubit gates (TQQGs) the entangling power can be compactly written in terms of a two-qubit gate local invariant\cite{EntanglingPowerSankara} and sets values  to classify  TQQGs as perfect entanglers\cite{SpecialPerfectEntanglersRezakhani,EntanglingPowerSankara}, that is to say, a quantum gate that at least produces a Bell state out of a separable state.  Nevertheless, these expressions do not quantify the entangling power of gates acting irreducibly on the symmetric subspace, called symmetric quantum gates (SQGs), because they involve a contribution from separable non-symmetric states. Here we will derive the appropriate expression of the entangling power for SQGs and find an onset value above which they can be classified as perfect entanglers. 

As noted by Zhang \textit{et al.}\cite{GeometricTheoryNonlocal}, distinct TQQGs can be put together into sets whose elements differ by local transformations, called \textit{local equivalence classes} (LECs). By group theoretical methods, the authors were able to represent all LECs of TQQGs classes on a tetrahedron, and showed that half of its volume is occupied by perfect entanglers. 
Motivated by Reference \onlinecite{GeometricTheoryNonlocal}, we develop a geometric description of the LECs of symmetric gates. We found that these are characterized by a periodic set of points on a plane, displaying hexagonal symmetry
with lattice vectors determined by the $\mathfrak{su}(3)$ algebra root vectors. This allows us to identify a minimum extension where all inequivalent LECs of SQGs are located, known as the \textit{Weyl chamber}, and to show that one fourth of it is occupied by perfect entanglers.  This geometric approach as well as the entangling power concept are relevant to study operations in general three-level systems.   

This paper is organized as follows. In sections II and III we
present brief descriptions of the Majorana representation and of the Cartan decomposition of TQQGs and how it is related to SQGs, respectively. In section IV the appropriate local-invariant for SQGs is defined. Section V is devoted to analyse the entangling power and use it to classify SQGs as perfect entanglers. In section VI the developed formalism is applied to some example models involving two interacting spins\cite{HeisenbergModelXYZ}, the three-level Lipkin-Meshkov-Glick model from nuclear physics\cite{LIPKIN}, and two coupled quantized oscillators through the cross-Kerr effect of quantum optics\cite{Battacharya}. Section VII is devoted to conclusions. Appendix A includes the derivation of the entangling power formula while appendix B contains a theorem which allows to classify SQGs as perfect entanglers.


\section{Majorana stellar representation}

The Majorana representation is a geometric depiction of quantum states contained in a finite Hilbert space, which can give insight into their entangling properties\cite{LiuBerryEntanglement}. The idea behind its construction is to obtain a complex polynomial out of the probability amplitudes that define a state for some fixed basis. The roots of such a polynomial can be represented by points in the Argand diagram, and mapped into a sphere by stereographic projection\cite{MajoranaAtomi,LiuBerryEntanglement}. 
For a quantum state $\ket{\psi} = \sum a_k\ket{k}$ in a 
$n$ dimensional Hilbert space, the Majorana polynomial is given by
\begin{align}
    \sum\limits_{k=1}^{n}&\frac{(-1)^ka_{k}}{\sqrt{(n-k)!k!}}z^{n-k} = 0.\label{MajoranaPolynomial}
\end{align}
\noindent The solutions $\{z_k\}$ lie on the complex plane, and their projection onto the Riemann sphere is made by the following association $z_k = \tan(\theta_k/2)e^{i\phi_k}$. Each root $z_k$ is called a Majorana star, while the whole set of roots is denoted as the Majorana constellation of the quantum state $\ket{\psi}$. In fact, any two quantum states with the same constellation are in fact equivalent, up to a global phase, which makes this representation a good description of their projective space.

The $n+1$ dimensional Hilbert space has a bijection onto the space of $n$ qubits with particle permutation symmetry. This implies that symmetric-qubit states have associated Majorana constellations. The $z_k$ roots define the components of a ket state in the symmetric space, by the relation
\begin{align}
    \ket{\psi} &= \frac{1}{\sqrt{n!}N_n}\sum\limits_{P}^{n}\ket{u_{P(1)},u_{P(2)},...,u_{P(n)}},
\end{align}
\noindent where $\ket{u_i}$ represents the 1-qubit state $(\cos\frac{\theta_i}{2}\ e^{i\phi_i}\sin\frac{\theta_i}{2})^T$; the $P$ symbol denotes all the possible permutations of the $\ket{u_i}$ states and $N_n$ is a normalization coefficient\cite{BerryPhaseMajorana}. Thus, the Majorana constellation serves as the mapping between an $n+1$ dimensional Hilbert space and the space of $n$ qubit symmetric wavefunctions.

For symmetric two-qubit states, the Majorana constellation consists of two stars. As shown in Reference \onlinecite{LiuBerryEntanglement}, their concurrence is proportional to the square of the chordal distance between the stars. This 
implies that states with zero concurrence have their Majorana stars on the same position, or equivalently said,  have degenerate stars. On the contrary, maximally entangled states have stars occupying antipodal positions on the sphere. For mono-partite three-level systems we may also speak of entangled states as those whose Majorana stars are not coincident. 

\begin{figure}[t!]
     \centering
     \includegraphics[scale=0.28]{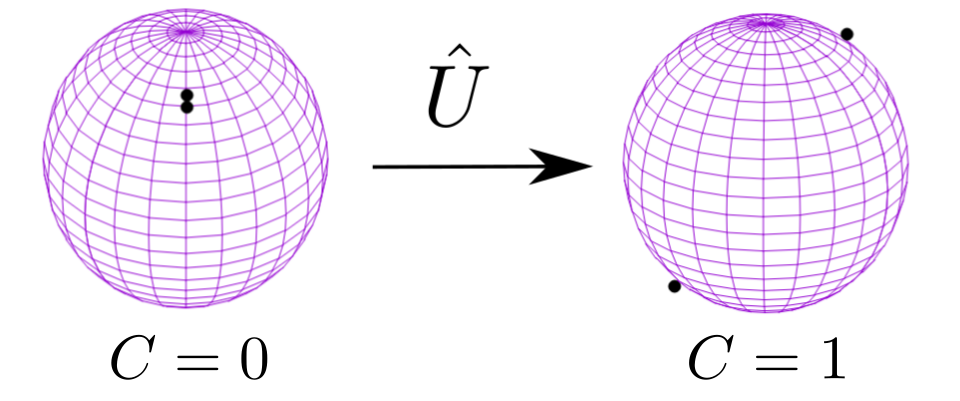}
     \caption{Action of a quantum gate on a separable initial state (with concurrence $C=0$) which ends up as a Bell state ($C = 1$) on the Majorana sphere.}
     \label{fig:fig1}
 \end{figure}

On the same line of thought, three-level transformations will be referred to as entangling as long as they can produce a state with non-degenerate Majorana stars from a state with a degenerate constellation. Fig.\,\ref{fig:fig1} illustrates a separable state (degenerate constellation) which, after being acted on by a SQG ends up as a fully entangled state (antipodal Majorana stars). Examples of these gates in three-level systems are phase gates and the $SWAP_{12}$ and $SWAP_{23}$, which have found applications in qutrit-based quantum computing\cite{Arvind}.



\section{Cartan decomposition of unitary two-qubit transformations  }\label{Sec:CartanDecomp}

The whole set of transformations on the Hilbert space  of two-qubits can be classified as local and non-local. Local operations are those physical processes that act separately only on one component of the bipartite system and, as a consequence, do not change the entanglement properties of the state. Local two-qubit gates can always be written as tensor product of one-qubit operations:
\begin{align}\label{eq:localU}
    \hat{V}^{(12)} &= \hat{V}^{(1)}\otimes\hat{V}^{(2)},
\end{align}
\noindent which belong to the $SU(2)\otimes SU(2)$ Lie group. We will restrict $\hat{V}$ to denote transformations on the two-qubit space, not necessarily symmetric. 

In general, TQQGs $\in SU(4)$ that cannot be written as in \eqref{eq:localU} are called \textit{non-local}. There is a very concise way of writing every element of $SU(4)$ given by the Cartan decomposition of the group. Namely, for every $\hat{V}\in SU(4)$, we have the following identity\cite{Knapp}
\begin{align}
    \hat{V} &= \hat{K}_1\hat{A}\hat{K}_2,\label{eq:DescG}\\
    \hat{A} &= \exp\left[\frac{i}{2}\sum\limits_k c_k\hat{\sigma}^{(1)}_{x_k}\otimes\hat{\sigma}^{(2)}_{x_k}\right],\label{eq:nonlocalA}
\end{align}
The $\hat{K}$ factors belong to the $SU(2)\otimes SU(2)$ Lie group, hence they are local. As usual, the $\hat{\sigma}_{x_k}$ operators denote the Pauli matrices, with $k=1,2,3$ and $x_k=x,y,z$. The $\hat{A}$ factor contains the non-local part of the quantum gates, and is given by the exponential of linear combinations of the operators $\hat{\sigma}^{(1)}_{x_i}\otimes\hat{\sigma}^{(2)}_{x_i}$. 
TQQGs that differ only by a $\hat{K}$ factor are said to be in the same \textit{local equivalence class}. These set of operators span the Cartan subalgebra of the $SU(4)$ Lie group, which is a maximally commuting subalgebra of $\mathfrak{su}(4)$\cite{GeometricTheoryNonlocal}. It is seen that the $(c_1,c_2,c_3)$ coordinates have a period of $\pi$ each, and thus the  topological structure of the local equivalence classes is a $3$-torus\cite{GeometricTheoryNonlocal}. The $\boldsymbol{c} = (c_1,c_2,c_3)$ point will be called {\it geometrical point} hereafter. For a more detailed discussion of the Cartan decomposition of $SU(4)$ and its algebra, namely the $\mathfrak{su}(4)$ Lie algebra, see Reference \onlinecite{GeometricTheoryNonlocal}.

There is a special case of TQQGs that act  irreducibly on the symmetric and anti-symmetric two-qubit ket spaces. Therefore, if $\hat{V}^{(r)}$ is an element of such a special set, in which $(r)$ denotes a reducible representation, it has a matrix form
\begin{align}\label{eq:SU(4)red}
    \hat{V}^{(r)} &= \left(\begin{matrix}
    \hat{U} & 0 \\
    0       & 1 \\
    \end{matrix} \right),
\end{align}
\noindent where $\hat{U}\in SU(3)$
acts on any symmetric linear combination of the computational basis; $\hat{U}$ is thus the SQG we are interested in. The last factor acts on the anti-symmetric Bell state $\ket{\phi^-}=\frac{1}{\sqrt{2}}(\ket{01}-\ket{10})$.

Since reducible gates are a subgroup of $SU(4)$, the Cartan decomposition holds for all elements of the form \eqref{eq:SU(4)red}. Also, for reducible TQQGs, the Cartan decomposition is composed of reducible factors. To see this, first let us note that $\hat{A}$ is reducible, as will be seen in the next section. With this, it is readily shown that a sufficient condition for $\hat{V}$ to be reducible is that $\hat{K}$ be reducible. To show that reducibility of $\hat{V}$ implies the reducibility of $\hat{K}$ consider the product
\begin{align}
    \left(\begin{matrix}
    \hat{U} & 0 \\
    0  &    1
    \end{matrix}\right) &= 
    \left(\begin{matrix}
    K^{(s)}_1 &  K'_1 \\
    K''_1     &  K^{(a)}_1
    \end{matrix}\right)
    \left(\begin{matrix}
    A^{(s)} &  0 \\
    0       &  A^{(a)}
    \end{matrix}\right)
    \left(\begin{matrix}
    K^{(s)}_2 &  K'_2 \\
    K''_2     & K^{(a)}_2
    \end{matrix}\right).\nonumber
\end{align}

\noindent The upper and lower off-diagonal elements (primed and doubled-primed) are three-dimensional column and row vectors, respectively. The $(s)$ and $(a)$ upper indices denote the $3\times 3$ matrix acting on the symmetric subspace and the scalar acting on the anti-symmetric subspace, severally. By explicit evaluation of the right hand side of the equation above, it is seen that $K'_1 = K'_2 = 0$ and  $K''_1 = K''_2 = 0$, which implies that the $\hat{K}$ factors are also reducible. The $\hat{K}$ factors for SQGs are forcefully of the form $e^{-i\frac{\theta}{2}\boldsymbol{\hat{n}}\cdot\boldsymbol{\hat{\sigma}}}\otimes e^{-i\frac{\theta}{2}\boldsymbol{\hat{n}}\cdot\boldsymbol{\hat{\sigma}}}$, with $\boldsymbol{\hat{n}}$ and $\theta$ a unit vector and a rotation angle. It can be shown that the $\hat{K}^{(s)}$ factors belong to the $SU(2)$ group in the spin-$1$ representation\cite{Jeevanjee}, which are seen as $SO(3)$ rotations on the Majorana sphere\cite{Arvind,LiuBerryEntanglement}.


\section{Local invariants and equivalence classes of symmetric quantum gates}\label{Sec:LocalInvariants}

 TQQGs that are equivalent up to a local gate factor (see eq.\eqref{eq:DescG}) have the same local invariants \cite{NonlocalPropertiesMakhlin,GeometricTheoryNonlocal}. Local invariants for two-qubit quantum gates are determined by the set of eigenvalues of
 \begin{align}\label{matrizm}
     \hat{M} &= \hat{V}^T_B\hat{V}_B,
 \end{align}
 
 \noindent where the label $B$ indicates that the gate $\hat{V}$ is expressed in the Bell basis $B=\{ \ket{\psi^+},\ket{\psi^-},\ket{\phi^+},\ket{\phi^-}\}$. 
 The transformation matrix between the Bell states and the computational basis, ordered as $\{\ket{00},\ket{01},\ket{10},\ket{11}\}$, is
 \begin{align}\label{eq:mapaBellComp}
     \hat{Q}^{\dagger} &= \frac{1}{\sqrt{2}}\left(\begin{matrix}
     1   &   0   &   0   &   1   \\
     0   &   i   &   i   &   0   \\
     i   &   0   &   0   &  -i   \\
     0   &   1   &  -1   &   0   \\
     \end{matrix}\right),
 \end{align}
 
 \noindent where the definition of the Bell states is evident from the matrix above; note that the last row corresponds to the anti-symmetric one. The Lie algebra of the local components of two-qubit quantum gate is isomorphic to the Lie algebra of the $SO(4)$ group\cite{GeometricTheoryNonlocal}, through the map defined by eq.\eqref{eq:mapaBellComp}.  Thus, any two-qubit quantum gate in the Bell basis, whose decomposition is given by \eqref{eq:DescG}, becomes
 \begin{align}\label{eq:TwoQubitGateBell}
     \hat{V}_B &= \hat{O}_1\hat{F}\hat{O}_2,
 \end{align}
 
 \noindent where $\hat{O}_{1,2} = \hat{Q}^{\dagger}\hat{K}_{1,2}\hat{Q}\in SO(4)$, and $\hat{F} = \hat{Q}^{\dagger}\hat{A}\hat{Q}$ which is diagonal in this basis. The Bell states 
 $\ket{\psi^+},\ket{\psi^-},\ket{\phi^+},\ket{\phi^-}$ are thus eigenstates of the $\hat{A}$ matrix, with respective eigenvalues 
\begin{subequations}\label{eq:EigenvalF}
 \begin{align}
     \lambda_1 &= e^{i(c_1-c_2+c_3)/2},\\
     \lambda_2 &= e^{i(c_1+c_2-c_3)/2},\\
     \lambda_3 &= e^{i(-c_1+c_2+c_3)/2},\\
     \lambda_4 &= e^{-i(c_1+c_2+c_3)/2}.
 \end{align}
\end{subequations}

The eigenvalues of the matrix $\hat{M}$ are determined by the quantities $(\Tr\,{\hat{M}})^2$ and $(\Tr\,{\hat{M}})^2-\Tr\,\hat{M}^{2}$, which in turn serve to define the local invariants of two qubit quantum gates, namely
\begin{subequations}
\begin{align}
    G_1 &= \frac{1}{16}(\Tr\,\hat{M})^2,\\
    G_2 &= \frac{1}{4}[(\Tr\,\hat{M})^2-\Tr\,\hat{M}^2] \ .
\end{align}
\end{subequations}

\noindent Thus, distinct TQQGs having the same local invariants are said to be locally equivalent.
 
 Reducible TQQGs can be expressed as in eq.\eqref{eq:SU(4)red} and their action on the symmetric subspace only depends on $\hat{U}$. The local invariant of the symmetric part of the gate is determined by the eigenvalues of
 \begin{align}\label{eq:symlocalm}
     \hat{m} &= \hat{U}^{T}_B\hat{U}_B,
 \end{align}
 
 \noindent and is then independent of the $\lambda_4$ eigenvalue.
 Since we have considered special unitary gates, this implies  
 \begin{align}\label{eq:planoseta}
 c_1+c_2+c_3 = 0.
 \end{align}
 
 \noindent Had we regarded general unitary gates, removing the extra phase factor would lead to the same condition above, thus all LECs can be located in the plane defined in \eqref{eq:planoseta}.
 
 \begin{figure}[t!]
     \centering
     \includegraphics[scale=0.45]{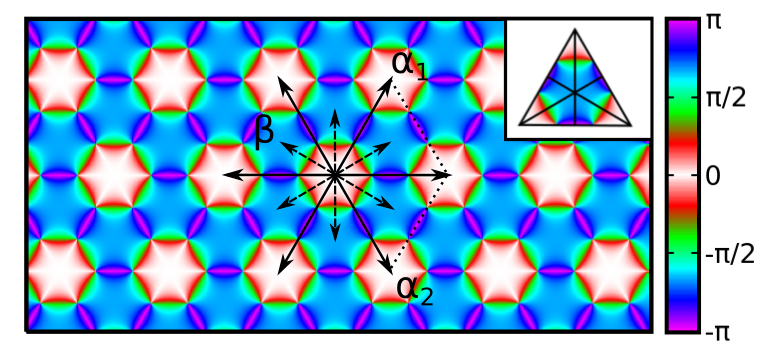}
     \caption{Phase plot of $\Tr\hat{m}$. The $\boldsymbol{\alpha}_1$ and $\boldsymbol{\alpha}_2$ are the root vectors of the $\mathfrak{su}(3)$ Lie algebra. 
     Every pair of antiparallel short arrows lies in a plane belonging to the set that generate the Weyl group. The inset shows a subcell divided into six slices, each one corresponding to a Weyl chamber.}
     \label{fig:WeylChamber}
 \end{figure}

The secular equation of the matrix $\hat{m}$ is given by
\begin{align}
    \lambda^3 - \Tr(\hat{m})\lambda^2-\Tr^*(\hat{m})\lambda -1 &= 0,
\end{align} 
where we have used \eqref{eq:TwoQubitGateBell} and \eqref{eq:EigenvalF} to simplify 
the related factor $\Tr^2(\hat{m})-\Tr(\hat{m}^2)$.
Thus, for SQGs the eigenvalues of $\hat{m}$ are determined by its trace. The argument of $\Tr\hat{m}$ is plotted in Fig.\ref{fig:WeylChamber}.

We define the SQG local invariant as
\begin{align}
    G &= \frac{1}{9}\left[\Tr(\hat{m})\right]^2.
\end{align}

\noindent  The norm of $G$ can be compactly written in terms of the $(c_1,c_2,c_3)$ vector as
\begin{align}\label{eq:symlocalinvG}
    |G| &= 1-\frac{4}{9}\left[\sin^2(c_{12})+\sin^2(c_{13})+\sin^2(c_{32})\right],
\end{align}

\noindent where $c_{ij}$ is a short-hand notation for $c_i-c_j$. Thus, distinct SQGs having the same value of $G$ are said to be locally equivalent.

The whole extension of the $O$-{\it plane} (\ref{eq:planoseta}) has more information than is actually needed, given the periodicity the matrix $\hat{A}$ (eq.\eqref{eq:nonlocalA}) up to local gate factors $\hat{K}$. Consider the vectors $\boldsymbol{\alpha}_1=(-\pi,0,\pi)$ and $\boldsymbol{\alpha}_2=(0,\pi,-\pi)$  lying on the $O$-plane. 
The $\hat{A}$ matrix is obviously periodic along these directions. Also, the angle between them is $2\pi/3$. Thus, SQGs whose geometrical point differ by a $n\boldsymbol{\alpha}_1+m\boldsymbol{\alpha}_2$ translation ($n,m\in\mathbb{Z}$) are locally equivalent, and the whole set of local equivalent classes can be found within the hexagonal primitive cell spanned by  $\boldsymbol{\alpha}_1$ and $\boldsymbol{\alpha}_2$, which is displayed in Fig.\,\ref{fig:WeylChamber} by the area between these vectors and the dotted lines. The vectors $\boldsymbol{\alpha}_1$ and $\boldsymbol{\alpha}_2$ are the root vectors of the $\mathfrak{su}(3)$ Lie algebra, and 
the set $\{\pm\boldsymbol{\alpha}_1,\pm\boldsymbol{\alpha}_2,\pm(\boldsymbol{\alpha}_1+\boldsymbol{\alpha}_2)\}$ (solid-arrows in Fig.\,\ref{fig:WeylChamber}) form the root space of the algebra\cite{HallLieAlgebras}.

Consider the reflection matrix $(\hat{\sigma}_{\boldsymbol{\hat{\beta}}})_{ij}=\delta_{ij}-2\beta_i\beta_j/\beta^2\ (i,j=x,y,z)$
on the plane normal to the unit vector $\boldsymbol{\hat{\beta}}$, which is obtained from
any vector lying between two successive root vectors (dotted-line arrows). 
It can be checked that the effect of $\hat{\sigma}_{\boldsymbol{\hat{\beta}}}$ is to permute and complex-conjugate the eigenvalues of the $\hat{m}$ matrix. 
From Fig.\,\ref{fig:WeylChamber} it is seen that reflection on the plane normal to the vertical $\boldsymbol{\hat{\beta}}$ vectors interchanges the triangles composing the unit cell depicted there. This means that knowledge of $\Tr\hat{m}$ on a subcell 
determines its value on the other subcell by complex conjugation, and thus all LECs can be located in just one subcell.

Reflection on the planes normal to the root vectors generates the Weyl group of $\mathfrak{su}(3)$; the corresponding reflection matrices are given by $\hat{\sigma}_{\boldsymbol{\hat{\alpha}}}$, where $\boldsymbol{\hat{\alpha}}$ is a normalized root vector. The action of this group on the $\hat{A}$ matrix is permuting its eigenvalues, and thus leaves the character of $\hat{m}$ invariant. As in the general two-qubit case\cite{GeometricTheoryNonlocal}, the Weyl group allows to define a minimum extension containing the whole set of local equivalence classes, called the \textit{Weyl chamber} \cite{GeometricTheoryNonlocal,HallLieAlgebras}. By bisecting one triangle of the primitive cell by the planes normal to the root space we get the inset of Fig.\,\ref{fig:WeylChamber}. Every slice of the triangle contains all the local-equivalence classes of SQGs up to complex conjugation and represents a Weyl chamber.  Thus, by means of the $\hat{\sigma}_{\boldsymbol{\hat{\beta}}}$ and $\hat{\sigma}_{\boldsymbol{\hat{\alpha}}}$ reflections, we have reduced to a minimum the extension needed to locate all distinct LECs. We will take advantage of this in the next section  to obtain the ratio of perfect entanglers to all the possible SQGs. 

\section{Entangling power of symmetric two-qubit quantum gates}
\subsection{Expression and properties}

 \begin{figure}
     \centering
     \includegraphics[scale=0.55]{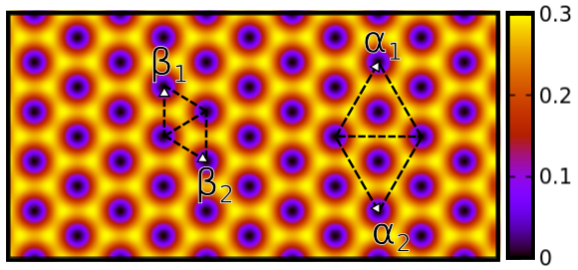}
     \caption{Entangling power $ep$ of SQGs on the $O$-plane. The parallelogram spanned by the root vectors $\boldsymbol{\alpha_1},\boldsymbol{\alpha_2}$ corresponds to the primitive cell. The smaller section defined by $\boldsymbol{\beta_1},\boldsymbol{\beta_2}$ contains all the possible values $ep$.}
     \label{fig:ep}
 \end{figure}

The entangling power of a quantum gate acting on a bipartite system is a measure of the ability for these gates to create entangled states  from the set of all bipartite separable states. In the general two-qubit case\cite{ZanardiEntanglingPOwer}, the entangling power of a gate $\hat{V}$ (\ref{eq:DescG}) is defined as the average of the linear entropy
of entanglement $E(\ket{\psi}) = 1-\Tr(\hat{\rho}^2_1)$,
over the set of all separable symmetric two-qubit states with a uniform probability,
\begin{align}
ep(\hat{A}) &= \overline{E(\hat{V}\ket{\psi_1}\otimes\ket{\psi_2})},
\end{align}
with the bar indicating such an average.
Note that we have written $ep$ as a function of $\hat{A}$, since local transformations do not change the entanglement of a quantum state. The entanglement power is very informative, given that it can be compactly expressed in terms of the two-qubit $|G_1|$ local-invariant and can be used as an indicator to whether a quantum gate is a perfect entangler or not\cite{EntanglingPowerSankara}. Nevertheless, for SQGs acting on symmetric states only, this expression of the entangling power is not adequate since it takes into account all two-qubit separable states, not necessarily symmetric. Hence, we need to restrict the entangling power definition to symmetric states in order to obtain the correct expression. 
Accordingly, we define the entangling power of SQGs as 
\begin{align}\label{epsymmetric}
    ep(\hat{U}) &= \overline{E(\hat{U}\ket{u}\otimes\ket{u})}.
\end{align}
 \noindent where $\ket{u}$ is a 1-qubit state, as in the section II. 
 
 We are now going to derive an explicit formula for this expression. First of all, let us consider a uniform distribution of initial $\ket{u,u}$ states, for which the Majorana constellation consists of two stars in the same position. Referring to appendix \ref{appendixA} for the details in the derivation, the entangling of SQGs is
 \begin{align}\label{eq:epsymExplicit}
     ep(\hat{U}) &= \frac{3}{10}\left( 1-|G|\right),
 \end{align}
 \noindent with $|G|$ given by expression \eqref{eq:symlocalinvG}.  It is remarkable that the entangling power obtained can be so compactly expressed in terms of the local invariant $|G|$, which only depends on the trace of matrix $\hat{m}$ (\ref{eq:symlocalm}). This result resembles that of the entangling power for general two-qubit gates\cite{EntanglingPowerSankara}, namely $ep = 2(1-|G_1|)/9$.

The function $ep(c_1,c_2,c_3)$ (\ref{eq:epsymExplicit}) presents minimum and maximum magnitudes at the geometrical points $\boldsymbol{c}=(0,0,0)$ and $(-\pi/3,0,\pi/3)$, which are zero and $3/10$, respectively.
As Fig.\,\ref{fig:ep} suggests, these extreme values
are also reached in additional points $\boldsymbol{c}$, obtained through symmetry operations which are translations along $\boldsymbol{\beta}$ vectors and $C_6$ rotations about the $(1,1,1)$ axis. Note also that $ep$ is invariant under translations along this same vector, which means that the
same pattern as that shown in Fig.\,\ref{fig:ep} is displayed
in planes parallel to the $O$-plane \eqref{eq:planoseta}.


 \subsection{Perfect entanglers}
 
 A TQQG $\hat{V}$ is a perfect entangler if it is capable of producing a fully entangled state from a separable one. The condition for this is that the convex hull of eigenvalues of the $\hat{M}$ matrix contains the origin in the $\boldsymbol{c}$-space\cite{GeometricTheoryNonlocal}. For SQGs  the same condition holds applied to the corresponding matrix $\hat{m}$. The proof of this goes along the same lines as in the general case \cite{GeometricTheoryNonlocal}; in order to make the paper more self-contained we sketch it in appendix B.
 \begin{figure}
     \centering
     \includegraphics[scale=0.6]{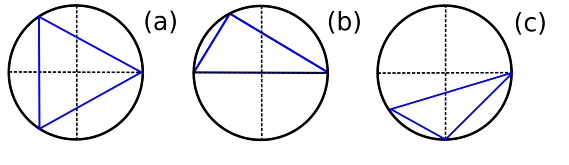}
     \caption{Convex hull of the matrix $\hat{m}$ associated to SQGs for several entanglement capabilities. The vertices on the unitary circle are defined by the phase of the eigenvalues of $\hat{m}$. (a) Perfect entangler with a maximum value of $ep$. (b) Perfect entangler with a minimum value of $ep$. (c)  Not a perfect entangler. The entangling character of the gate is geometrically determined by the location of the origin, inside (perfect) or outside (non-perfect) of the convex hull.}
     \label{fig:convexHull}
 \end{figure}
 
 Fig.\,\ref{fig:convexHull} shows the convex hulls of eigenvalues for three distinct cases. The eigenvalues of unitary matrices all have unit length, and thus the circle in the figure is unitary. The vertices of the triangles are defined by the phase of $\lambda_i^2$, which are the eigenvalues of $\hat{m}$. In (a) the eigenvalues are separated by $2\pi/3$. This case corresponds to the maximum value $ep = 3/10$ of the entangling power, since $\lambda^2_1 + \lambda^2_2 + \lambda^2_3=0$ and $|G| = 0$. 
 The case in (b) also represents a convex hull for a perfect entangling SQG, with the requirement that $\lambda^2_i = -\lambda^2_j$. This makes $|G| = 1/9$ and, as a consequence,the entangling power is then $4/15$. In fact, the $SWAP_{12}$ ($SWAP_{23}$) gate\cite{Arvind} have this value of $ep$, hence all SQGs with the same entangling power are locally equivalent to the $SWAP_{12}$. Since any deformation of such a convex hull such that it no longer contains the origin makes $ep$ less than $4/15$, this values is the minimum such that the corresponding SQGs are perfect entanglers.  A SQG with this convex hull can be built with coefficients  $(0,\pi/2,\pi/2)$. Other coefficients $(c_1,c_2,c_3)$ satisfying this condition can be obtained through symmetry transformations on the geometric point just given.  The case when the convex hull does not contain the origin is depicted in (c), the SQGs not being perfect entanglers ($ep<4/15$).
 
\begin{figure}
     \centering
     \includegraphics[scale = 0.3]{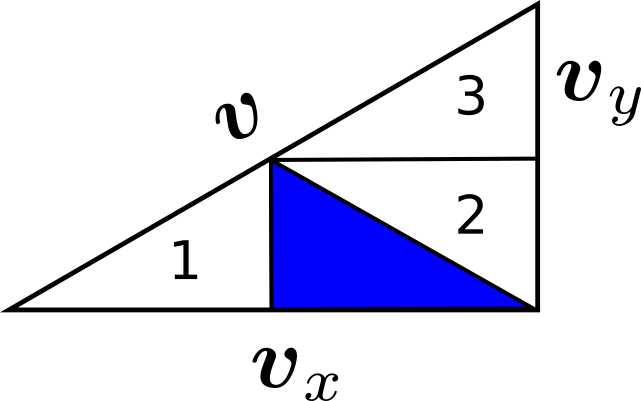}
     \caption{Weyl chamber of local-equivalence classes. The blue area contains all the perfect entanglers, and it is one fourth of the triangle.}
     \label{fig:PortionPE}
 \end{figure}

Now that we can classify SQGs as perfect entanglers or not according the geometrical point $\boldsymbol{c}$, we are at a position to calculate the ratio of perfect entanglers to non-perfect entanglers. To do this we will calculate the area on the Weyl chamber whose geometrical points correspond to perfect entangling SQGs. Let us restrict the $(c_1,c_2,c_3)$ to the Weyl chamber, as shown on Fig.\,\ref{fig:PortionPE}. Comparison with Fig.\,\ref{fig:ep} indicates the following association: $\boldsymbol{v} = \boldsymbol{\beta}_1+\boldsymbol{\beta}_2$, $\boldsymbol{v}_x = (\boldsymbol{\alpha}_1+\boldsymbol{\alpha}_2)/2$ and $\boldsymbol{v}_y = (\boldsymbol{\alpha}_1-\boldsymbol{\alpha}_2)/6$.  
As before, the $\boldsymbol{\alpha}_1$ and $\boldsymbol{\alpha}_2$ root vectors are taken as $\pi(-1,0,1)$ and $\pi(0,1,-1)$. Hence, the vectors on Fig.\,\ref{fig:PortionPE} are
$\boldsymbol{v}=\frac{\pi}{3}(-2,1,1)$, $\boldsymbol{v}_x = \frac{\pi}{2}(-1,1,0)$, and $\boldsymbol{v}_y = \frac{\pi}{6}(-1,-1,2)$.
 Any geometrical point on the Weyl chamber shown in Fig.\,\ref{fig:PortionPE} can be expressed as $\boldsymbol{c} = s_1\boldsymbol{v}_x+s_2\boldsymbol{v}_y$, where $s_1,s_2 \in [0,1]$ and $s_2\leq s_1$.  Upon multiplying the $\lambda^2$ eigenvalues by a total phase, its convex hull gets rotated, and does not affect whether or not it contains the origin. Thus, regarding $c_1+c_2+c_3 = 0$ and setting $\lambda^2_2 = 0$ (see eqs.\eqref{eq:EigenvalF}), we have
\begin{align}
    \phi_1 &= \pi(s_1+s_2),\label{eq:phi1}\\
    \phi_3 &= \pi(s_2-s_1),\label{eq:phi3}
\end{align}

\noindent where $\phi_1$ and $\phi_3$ are the phase angles of $\lambda^2_1$ and $\lambda^2_3$, respectively. With this, a SQG is a perfect entangler if and only if the following condition holds:
\begin{align}
     0 \leq \phi_1 \leq \pi \hspace{0.5cm} AND \hspace{0.5cm} 
    -\pi\leq  \phi_3 \leq -\pi+\phi_1,\label{eq:1rades}
\end{align}

\noindent where all phases are equivalent modulo $2\pi$. The case $\phi_1>\pi$ always yields non-perfect entanglers by the imposed conditions on $s_1$ and $s_2$ (see the discussion below). Let us analyse  all sections of the Weyl chamber to determine whether they are composed of perfect entanglers or not.

\textit{Area 1}. This area is constrained to the $(s_1,s_2)$ coordinates: $0\leq s_1 < 1/2$ and $0\leq s_2\leq s_1$. These inequalities imply that $s_1+s_2<1$, hence $0\leq\phi_1<\pi$, leaving us in the domain of \eqref{eq:1rades}. By substituting $\phi_3$ into the right side of \eqref{eq:1rades} we get $s_1 \geq 1/2$, which is a contradiction given the imposed conditions on $s_1$ and $s_2$. Thus, all geometrical points in this section of the Weyl chamber do not correspond to perfect entanglers.

\textit{Areas 2 and 3}.  For these regions we have the following restriction on the $(s_1,s_2)$ coordinates: $1/2< s_1\leq 1$ and $1- s_1 < s_2\leq s_1$. These imply $\phi_1>\pi$ and  $-\pi<\phi_3 < 0$. Thus, both $\lambda^2_1$ and $\lambda^2_3$ are on the lower half of the unit circle (see Fig.\ref{fig:convexHull}.c), none of them at $\pi$. The convex hull does not contain the origin and the geometrical points do not correspond to perfect entanglers.

\textit{Blue area}. In this case, the $(s_1,s_2)$ coordinates are constrained by the inequalities: $1/2\leq s_1\leq 1$ and $0\leq s_2 \leq 1-s_1$. This implies $\pi\geq \phi_1 \geq \pi/2$ and consequently we must focus on expression \eqref{eq:1rades}. Inserting eqs.\eqref{eq:phi1} and \eqref{eq:phi3} into the right inequality of \eqref{eq:1rades} we obtain $-1\leq s_2-s_1 \leq -1+s_1+s_2$. $-1\leq s_2-s_1 $ holds trivially, while the right hand side of the last inequality implies $s_1\geq 1-s_1$, which holds since we are considering $1\geq s_1 \geq 1/2$. Expression \eqref{eq:1rades} holds in this case and all the geometrical points contained in the blue area correspond to perfect entanglers. This section occupies one fourth of the Weyl chamber. For this reason, the perfect entanglers are one fourth of the total SQGs.

It is worth noting at this point that the geometric picture of LECs of SQGs does not trivially arises from that of TQQGs. For example, the Weyl Chamber of TQQGs is a tetrahedron that, without loss of generality, has one vertex on the origin $O$. One of its edges, called the $OA_3$ edge\cite{GeometricTheoryNonlocal,CharacterizingGeometricalEdges}, along geometrical points of the form $(c,c,c)$, contains a  point $P$ which represents a perfect entangler. Nevertheless, for SQGs, the convex hull on any geometrical point of the form $(c,c,c)$ does not contain the origin, as can be seen from inspection of eqs.\eqref{eq:EigenvalF} and, as a result, the entangling power is zero along the $OA_3$ edge. This fundamental difference between the geometry of LECs of SQGs and TQQGs proves that the former is not just a trivial special case of the latter.

\section{Examples}

In this section we will apply the theory developed so far to three distinct physical models with three-dimensional Hilbert spaces. We calculate the entangling power as a function of an independent parameter for the three models and find conditions on them to obtain perfect entanglers. The linear entropy on the Majorana sphere is computed, where each point on it corresponds to a state with degenerate Majorana constellation and the color indicates the value of the entanglement measure of the final state. We comment on some features of the spacial distribution of entanglement on the Majorana sphere.

\subsection{Anisotropic Heisenberg model with no cross-terms}

The anisotropic Heisenberg model of two interacting spins is represented by the Hamiltonian\cite{HeisenbergModelXYZ}
\begin{align}\label{eq:HeisenbergHam}
\hat{H}_H &= -\frac{1}{2}\left(I_x\hat{\sigma}^{(1)}_{x} \hat{\sigma}^{(2)}_{x}+I_y\hat{\sigma}^{(1)}_{y} \hat{\sigma}^{(2)}_{y}+I_z\hat{\sigma}^{(1)}_{z} \hat{\sigma}^{(2)}_{z}\right),
\end{align}

\noindent where $I_i$ are the spin coupling constants. This Hamiltonian is composed of the Cartan subalgebra elements of $\mathfrak{su}(4)$, and hence has a reducible representation into symmetric and anti-symmetric subspaces. The symmetric part of the time evolution operator is, in the Bell basis,
\begin{align}
    \hat{U}_H &= \left(\begin{matrix}
    e^{i(I_x-I_y+I_z)t/2} &   0     &   0   \\
    0        & e^{i(I_x+I_y-I_z)t/2}&   0   \\
    0        &  0       &  e^{i(-I_x+I_y+I_z)t/2}
    \end{matrix}\right).
\end{align}

From eq.\eqref{eq:HeisenbergHam}, the $\boldsymbol{c}$ coordinate vector is $(I_x,I_y,I_z)$. The entangling power becomes
\begin{align}
    ep &= \frac{2}{15}\left[\sin^2(I_{xy}t)+\sin^2(I_{yz}t) +\sin^2(I_{xz}t) \right],
\end{align}

\noindent where $I_{ij}=I_i-I_j$. 
Note that for equal spin coupling constants the entangling power is zero, which means that the isotropic Heisenberg model does not have any entangling power on the symmetric two-qubit subspace.  

\begin{figure}
    \centering
    \includegraphics[scale=0.45]{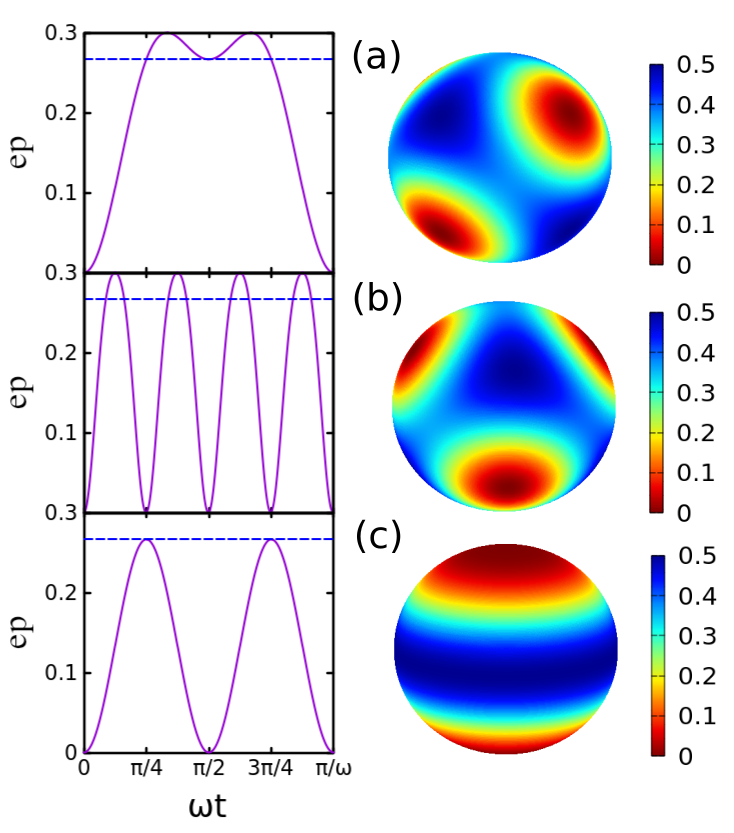}
    \caption{Left: entangling power for (a) the anisotropic Heisenberg model, (b) the Lipkin-Meshkov-Glick model and (c) the cross-Kerr interaction. The horizontal-dashed line indicates the lower bound for perfect entanglers ($ep = 4/15$). Right: Linear entropy of the final states obtained by applying $\hat{U}$ to the initial states with degenerate Majorana stars.}
    \label{fig:modelsEP}
\end{figure}

 Fig.\,\ref{fig:modelsEP}(a) shows the entangling power  as a function of $\omega t$ for the choice of parameters $I_y=0,I_x = -I_z = \omega$ (left panel). The maximum values are located at $\omega t = \pi/3, 2\pi/3$, as one would expect, while the minimum $ep$ for which the quantum gate is a perfect entangler is located at $\omega t = \pi/2$. 

The sphere on the right shows the linear entropy on the Majorana sphere; we have chosen $\omega t = \pi/3$ in order to obtain maximum entanglement. There are zones on which the quantum gate does not produce entanglement (red spots), and zones where the initial states become symmetric Bell states (blue spots). Even though it is not fully depicted here, there are exactly four low-entanglement zones and four high-entanglement zones, which form a tetrahedron on the sphere. As far as we have numerically checked, this tetrahedron distribution of entanglement on the Majorana sphere is a general feature of SQGs with maximum entangling power.

\subsection{Lipkin-Meshkov-Glick model}

The Lipkin-Meshkov-Glick model was firstly proposed to study the many body problem in nuclear physics\cite{LIPKIN}, and has also been useful to model the physics of molecular solids\cite{Garg_1993,HirschLMG} and critical phenomena in Bose-Einstein condensates\cite{ThermoLimitLMG}. The Hamiltonian describing the interaction is given by
\begin{align}
    \hat{H}_{L} &=  B\hat{J}_z + g_1\hat{J}^2_z-g_2\hat{J}^2_x.
\end{align}

This Hamiltonian commutes with the total angular momentum operator $\boldsymbol{J}^2$, which allows us to fix the $j$ value to unity, for which the model belongs to the $\mathfrak{su}(3)$ Lie algebra. This model can be written in matrix form as
\begin{align}
    \hat{H}_L &= \left(\begin{matrix}
    B+g_1 -g_2/2&   0     &   -g_2/2   \\
    0        & -g_2 &   0   \\
    -g_2/2        &  0       &  -B+g_1-g_2/2
    \end{matrix}\right).
\end{align}

\noindent The eigenvalues oh $\hat{H}_L$ are: $\lambda_0 = -g_2$ and $\lambda_\pm = g_1 - g_2/2 \pm \sqrt{B^2+g^2_2/4}$. The corresponding quantum gates of this system are given by $\exp{(-i\hat{H}_Lt)}$.

This model is a three-level system, and as such can be mapped to the symmetric two-qubit space. The isomorphism is given by mapping the angular momentum kets $\ket{1,1},\ket{1,0}$ and $\ket{1,-1}$ to the symmetric states $\ket{0,0},\frac{1}{\sqrt{2}}(\ket{0,1}+\ket{1,0}),\ket{11}$, respectively. With this, the transformation matrix from the spin-$1$ angular momentum basis to the symmetric Bell-basis is given by
\begin{align}
  \hat{T} &=  \left(\begin{matrix}
    1 & 0 & 1\\
    0 & \frac{i}{\sqrt{2}} & 0\\
    i  &  0   &  -i
    \end{matrix}\right).\label{eq:TransAngMomketsSymBell}
\end{align}

\noindent The definition of the Bell states should be evident from the matrix above. With this, the absolute square of the local invariant G, is given by
\begin{align}\label{eq:GL}
|G_L| &= \frac{1}{9}[1+4G_1\cos(2(g_1+g_2/2 )t) + 4G^2_1] \ .
\end{align}
where
\begin{align*}
    G_1 &= 1-\frac{g^2_2\sin^2(Rt)}{2R},\\
    R &= \sqrt{B^2+g^2_2/4}.
\end{align*}

\noindent The entangling power is readily obtained through eq.\eqref{eq:epsymExplicit}.

Fig.\,\ref{fig:modelsEP}(b) shows the entangling power for the choice of parameters $-2B/7 = g_1/2 = g_2/4 = \omega$, where $\omega$ is a fixed frequency. The entangling power displays an oscillating behaviour where the maximum $ep$ is reached at values of $\omega t$ closed to $\frac{\pi}{4}(n+1/2)$ for some integer $n\geq0$. On the right panel the linear entropy on the Majorana sphere for $\omega t = \pi/8$ is plotted. The colors on the sphere follow the tetrahedron-like patterns, as we have numerically checked for SQGs with maximum entangling power.

\subsection{Cross-Kerr Interaction}

The non-linear cross-Kerr effect in quantum optics is model by the Hamiltonian\cite{Battacharya}
\begin{align}
    \hat{H}_{CK} &= \omega_a\hat{a}^{\dagger}\hat{a}+\omega_b\hat{b}^{\dagger}\hat{b}+\omega_{CK}\hat{a}^{\dagger}\hat{a}\hat{b}^{\dagger}\hat{b}.
\end{align}

\noindent This Hamiltonian conserves the total number of excitations $\hat{N} =( \hat{a}^{\dagger}\hat{a}+\hat{b}^{\dagger}\hat{b})$. By transforming the Hamiltonian via the Schwinger operators of angular momentum\cite{Sakurai}, it is easily seen that $\hat{N}/2$ equals the total angular momentum $j$ of the resulting Hamiltonian. Let us work in the $j = 1$ representation. The cross-Kerr Hamiltonian becomes
\begin{align*}
    \hat{H}_{CK} &=  (\omega_a-\omega_b)\hat{J}_z - g_{CK}\hat{J}^2_z,
\end{align*}

The cross-Kerr interaction can be modeled as a Lipkin-Meshkov-Glick model after the identification $B = \omega_a-\omega_b, g_{1} = -g_{CK}, g_2 = 0$. The entangling power is then easily obtained from  eqs. \eqref{eq:GL} and \eqref{eq:epsymExplicit} and the transformation matrix \eqref{eq:TransAngMomketsSymBell}, 
\begin{align}
ep_{CK} &= \frac{4}{15}\sin^2(g_{CK}t).\nonumber
\end{align}

 Fig.\,\ref{fig:modelsEP}(c) shows the entangling power for a choice of parameters $B = g_1/2 = \omega$. The entangling power has an oscillating behaviour that never reaches the maximum possible value. The eigenvalues of the $\hat{m}$ matrix are $e^{-ig_{CK}t}, e^{-ig_{CK}t}, 1$. There is a pair of degenerate eigenvalues, and thus the convex hull contains the origin only when the vertices are on antipodal positions on the unit circle, which gives the minimum $ep$ for which $\hat{U}_{CK}$ is a perfect entangler. The linear entropy on the Majorana sphere is depicted on the right, which is very different from the spheres on panels (a) and (b). The antipodal high entangling spots on the sphere with a low entangling zone between them is a general feature of quantum gates with $ep = 4/15$, as we have seen from distinct numerical calculations, without regards on the details of the Hamiltonian.
 
 \section{Conclusions}

We have given a geometric classification of the LECs of SQGs, which turns out to be a plane with hexagonal symmetry. There we have identified the Weyl chamber, which is the minimum area that contains all distinct LECs. This geometric description contrasts with the general two-qubit case, for which the geometry is three-dimensional. The entangling power for SQGs is obtained in terms of the local-invariant. As was done in the general two-qubit case in reference \cite{GeometricTheoryNonlocal}, we give conditions for which a SQG is a perfect entangler, and have found that the perfect entanglers are 1/4 of all possible SQGs. Along this line, it is also found that perfect entanglers must have $ep\geq 4/15$. It is stressed that LECs of SQGs do not arise trivially from those of TQQGs, since there exist geometrical points corresponding to perfect entanglers in the latter case (TQQGs) that yield gates with zero $ep$ in the former (SQGs). The theoretic framework just developed can be applied to any three-level system, despite whether being bi-partite or not. The entangling power then refers to the capability of quantum gates to perform transformations that do not act as $SO(3)$ rotations on the Majorana constellation as, for example, some phase gates, $SWAP_{12}$ and $SWAP_{23}$ gates\cite{Arvind}. Finally, the theory just developed was applied to three physical models, namely, the anisotropic Heisenberg Model\cite{HeisenbergModelXYZ}, the Lipkin-Meshkov-Glick model\cite{LIPKIN} and two coupled quantized oscillators with cross-Kerr interaction\cite{Battacharya} and found some conditions on the Hamiltonian parameters to generate perfect entangling SQGs. 
Additional examples might include solid-state systems, or optical analogues, like assemblies of three quantum dots with few electrons used to study coherent control of quantum states\cite{BRANDES2005315}, Landau-Zener-St\"uckelberg interferometry\cite{LandauZenerPlatero}
or quantum transport \cite{TripleQuantumDot}, among other properties.

\section{Acknowledgements}

D.M.G. acknowledges financial support from CONACyT (M\'exico).
\medskip

\appendix
\section{Derivation of the entangling power for SQGs}\label{appendixA}

To derive an explicit expression for the entangling power of SQGs we first compute $\Tr[\hat{\rho}^2_1]$ and then integrate it over the unit sphere. The pure state density matrix obtained after applying a SQG $\hat{U}$ to a symmetric two-qubit separable state $\ket{u,u}$ is given by $\hat{U}\ket{u,u}\bra{u,u}\hat{U}^{\dagger}$. Since we are interested in this quantity in order to calculate the entropy of entanglement, the local contributions to the quantum gate $\hat{U}$ can be omitted, hence
\begin{align*}
    \hat{\rho} &= \hat{A}\ket{u,u}\bra{u,u}\hat{A}^{\dagger}.
\end{align*}
\noindent Now, by expressing $\ket{u,u}$ in the Bell basis through the matrix $\hat{Q}^{\dagger}$ (\ref{eq:mapaBellComp}), we find
\begin{align*}
    \hat{Q}^{\dagger}&\ket{u,u} = \frac{1}{\sqrt{2}}\left(\begin{matrix}
    a & ib & -ic & 0
    \end{matrix}\right)^T,
\end{align*}
\noindent where $a =  \cos\phi-i\sin\phi\cos\theta$, $b =  i\sin\theta$ and $c = -i\sin\phi+\cos\phi\cos\theta$. By transforming the $\hat{A}$ matrix to the Bell basis too, and after some manipulations we get the density matrix
\begin{align*}
    \hat{\rho} &= \frac{1}{4}\left( \begin{matrix}
  |A|^2 & AB^*  & AB^*  & AC^*  \\
  A^*B  & |B|^2 & |B|^2 & BC^*  \\
  A^*B  & |B|^2 & |B|^2 & BC^*  \\
  A^*C  & BC^*  & BC^*  & |C|^2 \\
 \end{matrix}\right),
\end{align*}
\noindent where the complex coefficients in the matrix elements are $A = \lambda_1a-i\lambda_3c$, $B = \lambda_2b$ and $C = \lambda_1a+i\lambda_3c$, with the $\lambda_i$ factors
 given in eqs.\,\eqref{eq:EigenvalF}. 
After reducing this matrix on one of the subsystem,
we obtain
\begin{align*}
    \Tr\hat{\rho}^2_1 &= \frac{1}{16}\left[ (|A|^2+|B|^2)^2+(|B|^2+|C|^2)^2 \right. \\&\hspace{3.5cm}\left.+ 2|AB^*+BC^*|^2 \right]. \nonumber
\end{align*}
What remains is to average this expression over all the possible symmetric separate states (degenerate Majorana constellations), where each state is equally likely to be obtained,
\begin{align*}
    \overline{\Tr\hat{\rho}^2_1} = \frac{1}{4\pi}\int_{\mathcal{A}} 
    \left[\Tr\hat{\rho}^2_1\right]\sin\theta\, d\theta d\phi,
\end{align*}
\noindent with integration performed over the surface $\mathcal{A}$
of the unit sphere. After a long but straightforward algebra, we obtain from (\ref{epsymmetric})
\begin{align*}
    ep &= \frac{2}{15}[\sin^2(c_{12})+\sin^2(c_{23})+\sin^2(c_{31})],
\end{align*}
which can be recast in the form (\ref{eq:epsymExplicit}) by using
the local invariant expression (\ref{eq:symlocalinvG}).


\section{Condition for a perfect entangler}\label{appendixB}

In order to determine the values of $ep$ for which a SQG is a perfect entangler, we 
need first a theorem based on the eigenvalues of the matrix $\hat{m}$. The proof runs very similar as that given in Reference \onlinecite{GeometricTheoryNonlocal} for the general two-qubit case, with the appropriate changes to focus on the symmetric case. First, we recall the definition of a convex hull.
 \begin{definition}
\textbf{(Convex hull)} The convex hull of $n+1$ points $\boldsymbol{p}_0,\boldsymbol{p}_1,...,\boldsymbol{p}_n \in \mathbb{R}^n$  is given by the set of all vectors of the form $\sum\limits^n_{i=0}\theta_i\boldsymbol{p}_i$,
where $\theta_i$ are non-negative real numbers satisfying $\sum\limits^n_{i=0}\theta_i=1$.
\end{definition}
\noindent Now, we state and prove the following theorem.

\begin{theorem}
 \textbf{(Perfect entanglers)} A symmetric two-qubit gate $\hat{U}$ is a perfect entangler if and only if the convex hull of eigenvalues of $\hat{m}$ contains zero.
\end{theorem}
\noindent \textit{Proof}: As explained in section \ref{Sec:CartanDecomp}, a general symmetric two-qubit gate can be decomposed as
$\hat{U} = \hat{K}^{(s)}_1\hat{A}^{(s)}\hat{K}^{(s)}_2$,
\noindent where $\hat{K}^{(s)}\in SU(2)$ in the spin-$1$ representation. Given a symmetric separable two-qubit state $\ket{\psi_o}=(a\ b\ b\ c)^T$, 
it is obtained for the concurrence\cite{EntangledSystems} 
of the state $\hat{U}\ket{\psi_o}$ that
$C(\hat{U}\ket{\psi_o})=
C(\hat{A}\ket{\psi_o})$,
\noindent and thus, for  $\hat{U}$ to be a perfect entangler, $\hat{A}$ has to be a perfect entangler;  we have also used $\hat{A}\ket{\psi_o} = \hat{A}^{(s)}\ket{\psi_o}$, since $\ket{\psi_o}$ is symmetric. Explicitly, 
$C(\hat{A}\ket{\psi_o})= \overline{\bra{\psi_o}}\hat{A}^{T}\hat{P}\hat{A}\ket{\psi_o}$,
where $\hat{P} = -\hat{\sigma}_y\otimes \hat{\sigma}_y$. This expression can be rewritten in terms of Bell states as
$C(\hat{A}\ket{\psi_o})=\overline{\bra{\psi_o}\hat{Q}})
\hat{F}^2(\hat{Q}^{\dagger}\ket{\psi_o}$, where we have used
the matrix $\hat{Q}$ (\ref{eq:mapaBellComp}) and the result
$\hat{Q}^T\hat{P}\hat{Q} = \mathds{1}$; the operator $\hat{F}$
is $\hat{Q}^{\dagger}\hat{A}\hat{Q}$.
Let $\hat{Q}^{\dagger}\ket{\psi_o} = \ket{\phi}$. Since $\ket{\psi_o}$ is a non-entangled state, we have
$C(\ket{\psi_o})= \overline{\bra{\psi_o}}\hat{P}\ket{\psi_o}=
 \overline{\bra{\phi}}\hat{Q}^T\hat{P}\hat{Q}\ket{\phi}=
\overline{\bra{\phi}} \mathds{1}\ket{\phi}=0$, which implies
\begin{align}\label{eq:ConcSeparable}
\phi^2_1+\phi^2_2+\phi^2_3=0,
\end{align}
\noindent where only three expansion coefficients appear, since $\ket{\psi_o}$ is a symmetric state, and thus has no projection on the antisymmetric axis. 
\noindent For $\hat{A}$ to be a perfect entangler, the concurrence of $\hat{A}\ket{\psi_o}$ must equal unity,  which along with the normalization condition yields $|\phi_1^2\lambda^2_1+\phi_2^2\lambda^2_2+\phi_3^2\lambda^2_3| = |\phi^2_1\lambda^2_1|+|\phi^2_2\lambda^2_2|+|\phi^2_3\lambda^2_3|$, where the eigenvalues $\lambda_i$ are given in eq.\eqref{eq:EigenvalF}. This equation holds if and only if there exists a number $\theta\in[0,2\pi)$, such that $\phi_j^2\lambda^2_j = |\phi_j|^2e^{i2\theta},\,j = 1,2,3,4$. With the abovementioned and eq.\eqref{eq:ConcSeparable}, we get
\begin{align*}
\phi^2_1 + \phi^2_2 + \phi^2_3 &= e^{i2\theta}\left( \frac{|\phi_1|^2}{\lambda^2_1} + \frac{|\phi_2|^2}{\lambda^2_2} + \frac{|\phi_3|^2}{\lambda^2_3} \right),\\
            &= e^{i2\theta}\left( |\phi_1|^2\overline{\lambda^2_1} + |\phi_2|^2\overline{\lambda^2_2} + |\phi_3|^2\overline{\lambda^2_3} \right),\\
            &= 0.
\end{align*}
\noindent By complex conjugation of the last equality, it follows that the convex hull of the eigenvalues of $m(\hat{U})$ contains the origin. The converse statement can be done following Reference \onlinecite{GeometricTheoryNonlocal}.

\bibliography{ref}

\begin{thebibliography}{29}%
\makeatletter
\providecommand \@ifxundefined [1]{%
 \@ifx{#1\undefined}
}%
\providecommand \@ifnum [1]{%
 \ifnum #1\expandafter \@firstoftwo
 \else \expandafter \@secondoftwo
 \fi
}%
\providecommand \@ifx [1]{%
 \ifx #1\expandafter \@firstoftwo
 \else \expandafter \@secondoftwo
 \fi
}%
\providecommand \natexlab [1]{#1}%
\providecommand \enquote  [1]{``#1''}%
\providecommand \bibnamefont  [1]{#1}%
\providecommand \bibfnamefont [1]{#1}%
\providecommand \citenamefont [1]{#1}%
\providecommand \href@noop [0]{\@secondoftwo}%
\providecommand \href [0]{\begingroup \@sanitize@url \@href}%
\providecommand \@href[1]{\@@startlink{#1}\@@href}%
\providecommand \@@href[1]{\endgroup#1\@@endlink}%
\providecommand \@sanitize@url [0]{\catcode `\\12\catcode `\$12\catcode
  `\&12\catcode `\#12\catcode `\^12\catcode `\_12\catcode `\%12\relax}%
\providecommand \@@startlink[1]{}%
\providecommand \@@endlink[0]{}%
\providecommand \url  [0]{\begingroup\@sanitize@url \@url }%
\providecommand \@url [1]{\endgroup\@href {#1}{\urlprefix }}%
\providecommand \urlprefix  [0]{URL }%
\providecommand \Eprint [0]{\href }%
\providecommand \doibase [0]{http://dx.doi.org/}%
\providecommand \selectlanguage [0]{\@gobble}%
\providecommand \bibinfo  [0]{\@secondoftwo}%
\providecommand \bibfield  [0]{\@secondoftwo}%
\providecommand \translation [1]{[#1]}%
\providecommand \BibitemOpen [0]{}%
\providecommand \bibitemStop [0]{}%
\providecommand \bibitemNoStop [0]{.\EOS\space}%
\providecommand \EOS [0]{\spacefactor3000\relax}%
\providecommand \BibitemShut  [1]{\csname bibitem#1\endcsname}%
\let\auto@bib@innerbib\@empty
\bibitem [{\citenamefont {Audrescht}(2007)}]{EntangledSystems}%
  \BibitemOpen
  \bibfield  {author} {\bibinfo {author} {\bibfnamefont {J.}~\bibnamefont
  {Audrescht}},\ }\enquote {\bibinfo {title} {Entanglement},}\ in\ \href@noop
  {} {\emph {\bibinfo {booktitle} {Entangled Systems}}}\ (\bibinfo  {publisher}
  {John Wiley \& Sons, Ltd},\ \bibinfo {year} {2007})\ Chap.~\bibinfo {chapter}
  {8}, pp.\ \bibinfo {pages} {143--168}\BibitemShut {NoStop}%
\bibitem [{\citenamefont {Horodecki}\ \emph {et~al.}(2009)\citenamefont
  {Horodecki}, \citenamefont {Horodecki}, \citenamefont {Horodecki},\ and\
  \citenamefont {Horodecki}}]{QuantumEntanglementHorodecki}%
  \BibitemOpen
  \bibfield  {author} {\bibinfo {author} {\bibfnamefont {R.}~\bibnamefont
  {Horodecki}}, \bibinfo {author} {\bibfnamefont {P.}~\bibnamefont
  {Horodecki}}, \bibinfo {author} {\bibfnamefont {M.}~\bibnamefont
  {Horodecki}}, \ and\ \bibinfo {author} {\bibfnamefont {K.}~\bibnamefont
  {Horodecki}},\ }\href {\doibase 10.1103/RevModPhys.81.865} {\bibfield
  {journal} {\bibinfo  {journal} {Rev. Mod. Phys.}\ }\textbf {\bibinfo {volume}
  {81}},\ \bibinfo {pages} {865} (\bibinfo {year} {2009})}\BibitemShut
  {NoStop}%
\bibitem [{\citenamefont {Muthukrishnan}\ and\ \citenamefont
  {Stroud}(2000)}]{MultivaluedLogicGates}%
  \BibitemOpen
  \bibfield  {author} {\bibinfo {author} {\bibfnamefont {A.}~\bibnamefont
  {Muthukrishnan}}\ and\ \bibinfo {author} {\bibfnamefont {C.~R.}\ \bibnamefont
  {Stroud}},\ }\href {\doibase 10.1103/PhysRevA.62.052309} {\bibfield
  {journal} {\bibinfo  {journal} {Phys. Rev. A}\ }\textbf {\bibinfo {volume}
  {62}},\ \bibinfo {pages} {052309} (\bibinfo {year} {2000})}\BibitemShut
  {NoStop}%
\bibitem [{\citenamefont {Jerger}\ \emph {et~al.}(2016)\citenamefont {Jerger},
  \citenamefont {Reshitnyk}, \citenamefont {Oppliger}, \citenamefont
  {Potočnik}, \citenamefont {Mondal}, \citenamefont {Wallraff}, \citenamefont
  {Goodenough}, \citenamefont {Wehner}, \citenamefont {Juliusson},
  \citenamefont {Langford},\ and\ \citenamefont
  {Fedorov}}]{Contextualitywithoutnonlocality}%
  \BibitemOpen
  \bibfield  {author} {\bibinfo {author} {\bibfnamefont {M.}~\bibnamefont
  {Jerger}}, \bibinfo {author} {\bibfnamefont {Y.}~\bibnamefont {Reshitnyk}},
  \bibinfo {author} {\bibfnamefont {M.}~\bibnamefont {Oppliger}}, \bibinfo
  {author} {\bibfnamefont {A.}~\bibnamefont {Potočnik}}, \bibinfo {author}
  {\bibfnamefont {M.}~\bibnamefont {Mondal}}, \bibinfo {author} {\bibfnamefont
  {A.}~\bibnamefont {Wallraff}}, \bibinfo {author} {\bibfnamefont
  {K.}~\bibnamefont {Goodenough}}, \bibinfo {author} {\bibfnamefont
  {S.}~\bibnamefont {Wehner}}, \bibinfo {author} {\bibfnamefont
  {K.}~\bibnamefont {Juliusson}}, \bibinfo {author} {\bibfnamefont {N.~K.}\
  \bibnamefont {Langford}}, \ and\ \bibinfo {author} {\bibfnamefont
  {A.}~\bibnamefont {Fedorov}},\ }\href {\doibase
  https://doi.org/10.1038/ncomms12930} {\bibfield  {journal} {\bibinfo
  {journal} {Nat. Commun.}\ }\textbf {\bibinfo {volume} {7}},\ \bibinfo {pages}
  {12930} (\bibinfo {year} {2016})}\BibitemShut {NoStop}%
\bibitem [{\citenamefont {Dogra}\ \emph {et~al.}(2018)\citenamefont {Dogra},
  \citenamefont {Dorai},\ and\ \citenamefont {Arvind}}]{Arvind}%
  \BibitemOpen
  \bibfield  {author} {\bibinfo {author} {\bibfnamefont {S.}~\bibnamefont
  {Dogra}}, \bibinfo {author} {\bibfnamefont {K.}~\bibnamefont {Dorai}}, \ and\
  \bibinfo {author} {\bibnamefont {Arvind}},\ }\href {\doibase
  10.1088/1361-6455/aaa69f} {\bibfield  {journal} {\bibinfo  {journal} {Journal
  of Physics B: Atomic, Molecular and Optical Physics}\ }\textbf {\bibinfo
  {volume} {51}},\ \bibinfo {pages} {045505} (\bibinfo {year}
  {2018})}\BibitemShut {NoStop}%
\bibitem [{\citenamefont {Randall}\ \emph {et~al.}(2018)\citenamefont
  {Randall}, \citenamefont {Lawrence}, \citenamefont {Webster}, \citenamefont
  {Weidt}, \citenamefont {Vitanov},\ and\ \citenamefont
  {Hensinger}}]{MultilevelSU(2)}%
  \BibitemOpen
  \bibfield  {author} {\bibinfo {author} {\bibfnamefont {J.}~\bibnamefont
  {Randall}}, \bibinfo {author} {\bibfnamefont {A.~M.}\ \bibnamefont
  {Lawrence}}, \bibinfo {author} {\bibfnamefont {S.~C.}\ \bibnamefont
  {Webster}}, \bibinfo {author} {\bibfnamefont {S.}~\bibnamefont {Weidt}},
  \bibinfo {author} {\bibfnamefont {N.~V.}\ \bibnamefont {Vitanov}}, \ and\
  \bibinfo {author} {\bibfnamefont {W.~K.}\ \bibnamefont {Hensinger}},\ }\href
  {\doibase 10.1103/PhysRevA.98.043414} {\bibfield  {journal} {\bibinfo
  {journal} {Phys. Rev. A}\ }\textbf {\bibinfo {volume} {98}},\ \bibinfo
  {pages} {043414} (\bibinfo {year} {2018})}\BibitemShut {NoStop}%
\bibitem [{\citenamefont {Majorana}(1932)}]{MajoranaAtomi}%
  \BibitemOpen
  \bibfield  {author} {\bibinfo {author} {\bibfnamefont {E.}~\bibnamefont
  {Majorana}},\ }\href@noop {} {\bibfield  {journal} {\bibinfo  {journal}
  {Nuovo Cim}\ }\textbf {\bibinfo {volume} {9}},\ \bibinfo {pages} {43}
  (\bibinfo {year} {1932})}\BibitemShut {NoStop}%
\bibitem [{\citenamefont {Devi}\ \emph {et~al.}(2012)\citenamefont {Devi},
  \citenamefont {Sudha},\ and\ \citenamefont {Rajagopal}}]{MajoranaMultiqubit}%
  \BibitemOpen
  \bibfield  {author} {\bibinfo {author} {\bibfnamefont {A.~R.~U.}\
  \bibnamefont {Devi}}, \bibinfo {author} {\bibnamefont {Sudha}}, \ and\
  \bibinfo {author} {\bibfnamefont {A.~K.}\ \bibnamefont {Rajagopal}},\ }\href
  {\doibase https://doi.org/10.1007/s11128-011-0280-8} {\bibfield  {journal}
  {\bibinfo  {journal} {Quantum Information Processing}\ }\textbf {\bibinfo
  {volume} {11}},\ \bibinfo {pages} {685} (\bibinfo {year} {2012})}\BibitemShut
  {NoStop}%
\bibitem [{\citenamefont {Liu}\ and\ \citenamefont
  {Fu}(2016)}]{LiuBerryEntanglement}%
  \BibitemOpen
  \bibfield  {author} {\bibinfo {author} {\bibfnamefont {H.~D.}\ \bibnamefont
  {Liu}}\ and\ \bibinfo {author} {\bibfnamefont {L.~B.}\ \bibnamefont {Fu}},\
  }\href {https://link.aps.org/doi/10.1103/PhysRevA.94.022123} {\bibfield
  {journal} {\bibinfo  {journal} {Phys. Rev. A}\ }\textbf {\bibinfo {volume}
  {94}},\ \bibinfo {pages} {022123} (\bibinfo {year} {2016})}\BibitemShut
  {NoStop}%
\bibitem [{\citenamefont {Zanardi}\ \emph {et~al.}(2000)\citenamefont
  {Zanardi}, \citenamefont {Zalka},\ and\ \citenamefont
  {Faoro}}]{ZanardiEntanglingPOwer}%
  \BibitemOpen
  \bibfield  {author} {\bibinfo {author} {\bibfnamefont {P.}~\bibnamefont
  {Zanardi}}, \bibinfo {author} {\bibfnamefont {C.}~\bibnamefont {Zalka}}, \
  and\ \bibinfo {author} {\bibfnamefont {L.}~\bibnamefont {Faoro}},\ }\href
  {\doibase 10.1103/PhysRevA.62.030301} {\bibfield  {journal} {\bibinfo
  {journal} {Phys. Rev. A}\ }\textbf {\bibinfo {volume} {62}},\ \bibinfo
  {pages} {030301} (\bibinfo {year} {2000})}\BibitemShut {NoStop}%
\bibitem [{\citenamefont {Balakrishnan}\ and\ \citenamefont
  {Sankaranarayanan}(2010)}]{EntanglingPowerSankara}%
  \BibitemOpen
  \bibfield  {author} {\bibinfo {author} {\bibfnamefont {S.}~\bibnamefont
  {Balakrishnan}}\ and\ \bibinfo {author} {\bibfnamefont {R.}~\bibnamefont
  {Sankaranarayanan}},\ }\href {\doibase 10.1103/PhysRevA.82.034301} {\bibfield
   {journal} {\bibinfo  {journal} {Phys. Rev. A}\ }\textbf {\bibinfo {volume}
  {82}},\ \bibinfo {pages} {034301} (\bibinfo {year} {2010})}\BibitemShut
  {NoStop}%
\bibitem [{\citenamefont
  {Rezakhani}(2004)}]{SpecialPerfectEntanglersRezakhani}%
  \BibitemOpen
  \bibfield  {author} {\bibinfo {author} {\bibfnamefont {A.~T.}\ \bibnamefont
  {Rezakhani}},\ }\href {\doibase 10.1103/PhysRevA.70.052313} {\bibfield
  {journal} {\bibinfo  {journal} {Phys. Rev. A}\ }\textbf {\bibinfo {volume}
  {70}},\ \bibinfo {pages} {052313} (\bibinfo {year} {2004})}\BibitemShut
  {NoStop}%
\bibitem [{\citenamefont {Zhang}\ \emph {et~al.}(2003)\citenamefont {Zhang},
  \citenamefont {Vala}, \citenamefont {Sastry},\ and\ \citenamefont
  {Whaley}}]{GeometricTheoryNonlocal}%
  \BibitemOpen
  \bibfield  {author} {\bibinfo {author} {\bibfnamefont {J.}~\bibnamefont
  {Zhang}}, \bibinfo {author} {\bibfnamefont {J.}~\bibnamefont {Vala}},
  \bibinfo {author} {\bibfnamefont {S.}~\bibnamefont {Sastry}}, \ and\ \bibinfo
  {author} {\bibfnamefont {K.~B.}\ \bibnamefont {Whaley}},\ }\href {\doibase
  10.1103/PhysRevA.67.042313} {\bibfield  {journal} {\bibinfo  {journal} {Phys.
  Rev. A}\ }\textbf {\bibinfo {volume} {67}},\ \bibinfo {pages} {042313}
  (\bibinfo {year} {2003})}\BibitemShut {NoStop}%
\bibitem [{\citenamefont {Abliz}\ \emph {et~al.}(2006)\citenamefont {Abliz},
  \citenamefont {Gao}, \citenamefont {Xie}, \citenamefont {Wu},\ and\
  \citenamefont {Liu}}]{HeisenbergModelXYZ}%
  \BibitemOpen
  \bibfield  {author} {\bibinfo {author} {\bibfnamefont {A.}~\bibnamefont
  {Abliz}}, \bibinfo {author} {\bibfnamefont {H.~J.}\ \bibnamefont {Gao}},
  \bibinfo {author} {\bibfnamefont {X.~C.}\ \bibnamefont {Xie}}, \bibinfo
  {author} {\bibfnamefont {Y.~S.}\ \bibnamefont {Wu}}, \ and\ \bibinfo {author}
  {\bibfnamefont {W.~M.}\ \bibnamefont {Liu}},\ }\href {\doibase
  10.1103/PhysRevA.74.052105} {\bibfield  {journal} {\bibinfo  {journal} {Phys.
  Rev. A}\ }\textbf {\bibinfo {volume} {74}},\ \bibinfo {pages} {052105}
  (\bibinfo {year} {2006})}\BibitemShut {NoStop}%
\bibitem [{\citenamefont {Lipkin}\ \emph {et~al.}(1965)\citenamefont {Lipkin},
  \citenamefont {Meshkov},\ and\ \citenamefont {Glick}}]{LIPKIN}%
  \BibitemOpen
  \bibfield  {author} {\bibinfo {author} {\bibfnamefont {H.}~\bibnamefont
  {Lipkin}}, \bibinfo {author} {\bibfnamefont {N.}~\bibnamefont {Meshkov}}, \
  and\ \bibinfo {author} {\bibfnamefont {A.}~\bibnamefont {Glick}},\ }\href
  {\doibase https://doi.org/10.1016/0029-5582(65)90862-X} {\bibfield  {journal}
  {\bibinfo  {journal} {Nuclear Physics}\ }\textbf {\bibinfo {volume} {62}},\
  \bibinfo {pages} {188} (\bibinfo {year} {1965})}\BibitemShut {NoStop}%
\bibitem [{\citenamefont {Bhattacharya}\ and\ \citenamefont
  {Shi}(2013)}]{Battacharya}%
  \BibitemOpen
  \bibfield  {author} {\bibinfo {author} {\bibfnamefont {M.}~\bibnamefont
  {Bhattacharya}}\ and\ \bibinfo {author} {\bibfnamefont {H.}~\bibnamefont
  {Shi}},\ }\href {\doibase https://doi.org/10.1119/1.4792696} {\bibfield
  {journal} {\bibinfo  {journal} {Am. J. Phys.}\ }\textbf {\bibinfo {volume}
  {81}},\ \bibinfo {pages} {267} (\bibinfo {year} {2013})}\BibitemShut
  {NoStop}%
\bibitem [{\citenamefont {Liu}\ and\ \citenamefont
  {Fu}(2014)}]{BerryPhaseMajorana}%
  \BibitemOpen
  \bibfield  {author} {\bibinfo {author} {\bibfnamefont {H.~D.}\ \bibnamefont
  {Liu}}\ and\ \bibinfo {author} {\bibfnamefont {L.~B.}\ \bibnamefont {Fu}},\
  }\href {https://link.aps.org/doi/10.1103/PhysRevLett.113.240403} {\bibfield
  {journal} {\bibinfo  {journal} {Phys. Rev. Lett.}\ }\textbf {\bibinfo
  {volume} {113}},\ \bibinfo {pages} {240403} (\bibinfo {year}
  {2014})}\BibitemShut {NoStop}%
\bibitem [{\citenamefont {Knapp}(1996)}]{Knapp}%
  \BibitemOpen
  \bibfield  {author} {\bibinfo {author} {\bibfnamefont {A.~W.}\ \bibnamefont
  {Knapp}},\ }\href@noop {} {\emph {\bibinfo {title} {Lie Groups Beyond an
  Introduction}}}\ (\bibinfo  {publisher} {Birkhäuser},\ \bibinfo {year}
  {1996})\BibitemShut {NoStop}%
\bibitem [{\citenamefont {{N. Jeevanjee}}(2015)}]{Jeevanjee}%
  \BibitemOpen
  \bibfield  {author} {\bibinfo {author} {\bibnamefont {{N. Jeevanjee}}},\
  }\href@noop {} {\emph {\bibinfo {title} {An Introduction to Tensors and Group
  Theory for Physicists.}}},\ \bibinfo {edition} {2nd}\ ed.\ (\bibinfo
  {publisher} {Birkh{\"a}user},\ \bibinfo {year} {2015})\BibitemShut {NoStop}%
\bibitem [{\citenamefont {Makhlin}(2002)}]{NonlocalPropertiesMakhlin}%
  \BibitemOpen
  \bibfield  {author} {\bibinfo {author} {\bibfnamefont {Y.}~\bibnamefont
  {Makhlin}},\ }\href {\doibase https://doi.org/10.1023/A:1022144002391}
  {\bibfield  {journal} {\bibinfo  {journal} {Quantum Information Processing}\
  }\textbf {\bibinfo {volume} {1}},\ \bibinfo {pages} {243} (\bibinfo {year}
  {2002})}\BibitemShut {NoStop}%
\bibitem [{\citenamefont {Hall}(2015)}]{HallLieAlgebras}%
  \BibitemOpen
  \bibfield  {author} {\bibinfo {author} {\bibfnamefont {B.}~\bibnamefont
  {Hall}},\ }\href {https://books.google.com.mx/books?id=v4vqjgEACAAJ} {\emph
  {\bibinfo {title} {Lie Groups, Lie Algebras and Representations: An
  Elementary Introduction}}}\ (\bibinfo  {publisher} {Springer International
  Publishing},\ \bibinfo {year} {2015})\BibitemShut {NoStop}%
\bibitem [{\citenamefont {Balakrishnan}\ and\ \citenamefont
  {Sankaranarayanan}(2009)}]{CharacterizingGeometricalEdges}%
  \BibitemOpen
  \bibfield  {author} {\bibinfo {author} {\bibfnamefont {S.}~\bibnamefont
  {Balakrishnan}}\ and\ \bibinfo {author} {\bibfnamefont {R.}~\bibnamefont
  {Sankaranarayanan}},\ }\href {\doibase 10.1103/PhysRevA.79.052339} {\bibfield
   {journal} {\bibinfo  {journal} {Phys. Rev. A}\ }\textbf {\bibinfo {volume}
  {79}},\ \bibinfo {pages} {052339} (\bibinfo {year} {2009})}\BibitemShut
  {NoStop}%
\bibitem [{\citenamefont {Garg}(1993)}]{Garg_1993}%
  \BibitemOpen
  \bibfield  {author} {\bibinfo {author} {\bibfnamefont {A.}~\bibnamefont
  {Garg}},\ }\href {\doibase 10.1209/0295-5075/22/3/008} {\bibfield  {journal}
  {\bibinfo  {journal} {Europhysics Letters ({EPL})}\ }\textbf {\bibinfo
  {volume} {22}},\ \bibinfo {pages} {205} (\bibinfo {year} {1993})}\BibitemShut
  {NoStop}%
\bibitem [{\citenamefont {Campos}\ and\ \citenamefont
  {Hirsch}(2011)}]{HirschLMG}%
  \BibitemOpen
  \bibfield  {author} {\bibinfo {author} {\bibfnamefont {J.}~\bibnamefont
  {Campos}}\ and\ \bibinfo {author} {\bibfnamefont {J.}~\bibnamefont
  {Hirsch}},\ }\href@noop {} {\bibfield  {journal} {\bibinfo  {journal} {Rev.
  Mex. Fis.}\ }\textbf {\bibinfo {volume} {57}},\ \bibinfo {pages} {56}
  (\bibinfo {year} {2011})}\BibitemShut {NoStop}%
\bibitem [{\citenamefont {Ribeiro}\ \emph {et~al.}(2007)\citenamefont
  {Ribeiro}, \citenamefont {Vidal},\ and\ \citenamefont
  {Mosseri}}]{ThermoLimitLMG}%
  \BibitemOpen
  \bibfield  {author} {\bibinfo {author} {\bibfnamefont {P.}~\bibnamefont
  {Ribeiro}}, \bibinfo {author} {\bibfnamefont {J.}~\bibnamefont {Vidal}}, \
  and\ \bibinfo {author} {\bibfnamefont {R.}~\bibnamefont {Mosseri}},\ }\href
  {\doibase 10.1103/PhysRevLett.99.050402} {\bibfield  {journal} {\bibinfo
  {journal} {Phys. Rev. Lett.}\ }\textbf {\bibinfo {volume} {99}},\ \bibinfo
  {pages} {050402} (\bibinfo {year} {2007})}\BibitemShut {NoStop}%
\bibitem [{\citenamefont {Sakurai}\ and\ \citenamefont
  {Napolitano}(1994)}]{Sakurai}%
  \BibitemOpen
  \bibfield  {author} {\bibinfo {author} {\bibfnamefont {J.}~\bibnamefont
  {Sakurai}}\ and\ \bibinfo {author} {\bibfnamefont {J.}~\bibnamefont
  {Napolitano}},\ }\href@noop {} {\emph {\bibinfo {title} {Modern Quantum
  Mechanics}}}\ (\bibinfo  {publisher} {Addison-Wesley Publishing Company},\
  \bibinfo {year} {1994})\BibitemShut {NoStop}%
\bibitem [{\citenamefont {Brandes}(2005)}]{BRANDES2005315}%
  \BibitemOpen
  \bibfield  {author} {\bibinfo {author} {\bibfnamefont {T.}~\bibnamefont
  {Brandes}},\ }\href {\doibase https://doi.org/10.1016/j.physrep.2004.12.002}
  {\bibfield  {journal} {\bibinfo  {journal} {Physics Reports}\ }\textbf
  {\bibinfo {volume} {408}},\ \bibinfo {pages} {315} (\bibinfo {year}
  {2005})}\BibitemShut {NoStop}%
\bibitem [{\citenamefont {Gallego-Marcos}\ \emph {et~al.}(2016)\citenamefont
  {Gallego-Marcos}, \citenamefont {S\'anchez},\ and\ \citenamefont
  {Platero}}]{LandauZenerPlatero}%
  \BibitemOpen
  \bibfield  {author} {\bibinfo {author} {\bibfnamefont {F.}~\bibnamefont
  {Gallego-Marcos}}, \bibinfo {author} {\bibfnamefont {R.}~\bibnamefont
  {S\'anchez}}, \ and\ \bibinfo {author} {\bibfnamefont {G.}~\bibnamefont
  {Platero}},\ }\href {\doibase 10.1103/PhysRevB.93.075424} {\bibfield
  {journal} {\bibinfo  {journal} {Phys. Rev. B}\ }\textbf {\bibinfo {volume}
  {93}},\ \bibinfo {pages} {075424} (\bibinfo {year} {2016})}\BibitemShut
  {NoStop}%
\bibitem [{\citenamefont {Maldonado}\ \emph {et~al.}(2018)\citenamefont
  {Maldonado}, \citenamefont {Villavicencio}, \citenamefont {Contreras-Pulido},
  \citenamefont {Cota},\ and\ \citenamefont {Maytorena}}]{TripleQuantumDot}%
  \BibitemOpen
  \bibfield  {author} {\bibinfo {author} {\bibfnamefont {I.}~\bibnamefont
  {Maldonado}}, \bibinfo {author} {\bibfnamefont {J.}~\bibnamefont
  {Villavicencio}}, \bibinfo {author} {\bibfnamefont {L.~D.}\ \bibnamefont
  {Contreras-Pulido}}, \bibinfo {author} {\bibfnamefont {E.}~\bibnamefont
  {Cota}}, \ and\ \bibinfo {author} {\bibfnamefont {J.~A.}\ \bibnamefont
  {Maytorena}},\ }\href {\doibase 10.1103/PhysRevB.97.195310} {\bibfield
  {journal} {\bibinfo  {journal} {Phys. Rev. B}\ }\textbf {\bibinfo {volume}
  {97}},\ \bibinfo {pages} {195310} (\bibinfo {year} {2018})}\BibitemShut
  {NoStop}%
\end{thebibliography}%

\end{document}